\documentclass[twocolumn,pre,a4paper]{revtex4}
\usepackage{epsfig}
\usepackage{amsfonts}
\usepackage[latin1]{inputenc}
\usepackage{natbib}
\def\dis{\displaystyle}
\begin{document} 
\title{On Translational Superfluidity and the Landau Criterion\\
 for Bose Gases in the Gross-Pitaevski Limit}

\author{Walter F. Wreszinski\footnote{Supported in part by CNPq.}} \affiliation{Departamento de F\'isica 
  Matem\'atica,Universidade de S\~ao Paulo, 
C.P. 66318 - 05315-970 -- S\~ao Paulo,
  Brazil}

\begin{abstract}
The two-fluid and Landau criteria for superfluidity are compared for trapped Bose gases. While the two-fluid criterion predicts translational superfluidity, it is suggested, on the basis of the homogeneous Gross-Pitaevski limit, that a necessary part of Landau's criterion, adequate for non-translationally invariant systems, does not hold for trapped Bose gases in the GP limit. As a consequence, if the compressibility is detected to be very large (infinite by experimental standards), the two-fluid criterion is seen to be the relevant one in case the system is a translational superfluid, while the Landau criterion is the relevant one if translational superfluidity is absent.


\vspace{0.2cm}

\end{abstract}

\maketitle
The experimental observation of Bose Einstein condensation (BEC) in a
vapor of Rb atoms at very low temperatures (of the order of $\,170
\times 10^{-9}K)\,$ in 1995 [1] challenged theoretical and
mathematical physicicists to find an explanation of the
phenomenon. Since atom-atom interactions are weak -- the range of
interaction was about $\, 10^{-6}$cm, while at the required densities
the interatomic spacing was about $\, 10^{-4}$cm-- the gas could be
considered to be almost perfect except for elastic collisions of hard
core type, and a theory was soon developed in terms of the so-called
Gross Pitaevski (G.P.) functional or equation [2] which may be
derived in the G.P. limit, defined by the condition:
\begin{equation}
\frac{Na}{L} = {\rm const.\/}
\end{equation}
where $a$ denotes the scattering length, $N\,$ is the number of particles
and $\, V=L^3\,$ is the volume enclosing the system. A mathematical
proof of BEC was only achieved in 2002 in a remarkable paper by Lieb and
Seiringer.~[3].

Setting $\,\mu = \hbar^2 /2m\,$, the Hamiltonian describing the system
may be written 
$$
H_N = - \mu \sum^N_{i=1}\limits H_{0i} + 
     \sum_{1\leq i < j\leq N}\limits 
       v(|\vec{x}_i - \vec{x}_j|)
\eqno{\rm{(2a)}}
$$
where $\,v\,$ is a {\em positive\/} potential satisfying certain
regularity conditions, and $\, H_0\,$ is the one-particle Hamiltonian 
$$
H_0 = \Delta  + V_{T} ( \vec{x}\,)
\eqno{\rm{(2b)}}
$$
acting on $\, L^2 (\Bbb{R}^3, d \vec{x}\,)\,$, with $\,\Delta\,$ the
Laplacian on $\,\Bbb{R}^3$. Above, $\, H_0\,$ describes
particles confined in a trap potential $\,V_{T}\,$ (such that
$\,V_{T}(\vec{x}\,) \rightarrow \infty\,$ as $\, |\vec{x}\,| \rightarrow
\infty$) (see, e.g., ref. [4]). In this letter,
we address the (not yet entirely clear) issue of superfluidity, most
particularly that of translational superfluidity.
In [5] the following criterion was used to prove superfluidity of these gases.
Let $E_0$ denote the ground state energy of the system in the rest frame and
$E^{'}_0$ the ground state energy, measured in the moving frame, when a 
velocity field $\vec{v}$ is imposed. Then for small $\vec{v}$
$$
E^{'}_0/N=E_0/N+\frac{\rho_s}{2m\rho}(\vec{v})^2+O(\vert \vec{v} \vert)^4
\eqno{\rm{(3)}}
$$
where $N$ is the particle number and $m$ the mass. The error term must be
bounded independently of $N$ (see [12], pg.44) and $\rho_s$ and $\rho$ denote, respectively, the density of the superfluid and the total particle density in the two-fluid picture (we shall call this ``the two-fluid criterion'' for brevity). By (4) the free Bose gas in the ground state is a superfluid, with 
$E_0=0,\rho_s=\rho$ and $E^{'}_0=1/2Nm(\vec{v})^2$ and no error term, while
it is not one by the Landau criterion, and thus the two criteria differ in
principle. The free Bose gas is not a physical system, but now that trapped 
Bose gases are an experimental reality, the fundamental question poses itself
a) do both the two-fluid and the Landau criterion predict translational 
superfluidity for these systems and b) if not, which criterion describes the 
physical property of superfluidity correctly? In order to address this question
we must now turn to a brief analysis of the Landau criterion. 

The original Landau criterion was formulated for homogeneous systems only,
for which momentum is conserved, and elementary excitations $\, \varepsilon
(\vec{k}\,)\,$ for fixed wave-vector $\, \vec{k}\,$ are defined. It asserts
that for drift velocities $\vec{v}$ such that
$|\vec v\,| \le v_c \equiv \inf{\varepsilon(\vec{k})/k}$ ,where $k=|\vec k\,|$
  , the flow is a superfluid, i.e., takes place without energy
dissipation[8]. For these systems the sound velocity $\, v_s\,$ in the medium is defined by
$$
v_s = \lim_{k \rightarrow 0}\limits 
  (\partial \varepsilon (k) / \partial k)
\eqno{\rm{(4)}}$$

A way to interpret and explain the Landau criterion is to view it as a \textbf{spectral condition} for a certain approximate Hamiltonian. This spectral aspect has played recently an important role in the explanation of superconductivity via superfluidity of the Cooper pairs in the BCS-BEC transition in traps [24]. At first one might be led to view the Landau condition as a (ground-state) stability condition for the Galilei-transformed Hamiltonian $H_{\vec{v},N}$ of $H_N$ given by (2a):
$$
H_{\vec{v},N} \equiv H_N+\vec{v}\cdot\vec{P_N}\ge0
\eqno{\rm{(2c)}}
$$
(as long as $|\vec v\,| \le v_c$), where $\vec{P_N}$ is the momentum operator: i.e., it is stability (positivity) of the Hamiltonian as viewed from the reference system attached to the pipe. Unfortunately, (2c) is not true ( see [22] and references given there), which may be interpreted as a sort of metastability of the Landau state (see also [15]). (2c) is, however, true (for $|\vec v\,| \le v_c$) when $H_N$ is taken as an approximating Hamiltonian describing a finite number (O(1)) of noninteracting elementary excitations, of the type of those analysed in [10b] of the forthcoming model (21) (see also [15]), or the hamiltonian of the weakly interacting Bose gas (WIBG) introduced in ref.[9]:
\begin{eqnarray*}
H_B\equiv \sum\varepsilon_ka^{*}_ka_k &+& \\
\sum\frac{\lambda_k}{2V}[a^{*}_0a_0(a^{*}_ka_k+a_{-k}a_{-k})] &+& \\
\sum\frac{\lambda_k}{2V}(a^{*}_ka^{*}_{-k}(a^{*}_0)^2)+\mbox{h.c.}
\end{eqnarray*}
$$\eqno{(2d)}$$

Above, $\varepsilon_k$ denotes the kinetic energy, $\lambda_k$ is related to the Fourier transform of the interaction potential, $V$ is the volume and $a$ are the usual Boson annihilation operators, related to a cubic box with periodic boundary conditions, and the sums exclude the point $k=0$ [9]. When the Bogoliubov c-number substitution $a_0/\sqrt{V}=\alpha_0$ is performed on (2d), $H_B$ becomes equivalent (after a Bogoliubov transformation) to a Hamiltonian of (an infinite number of) independent elementary excitations, which satisfies (2c) [9]. It should be emphasized that (2d) may be controlled in several ways, both with and without the c-number substitution [9].

The explanation of the stability (2c) for a Hamiltonian describing independent elementary excitations is that it describes, of course, the idealized situation in scattering theory, in which the excitations are infinitely far apart and do not interact. Instabilities may be brought about by their interaction: this physical picture has been made rigorous for spin-waves in the infinite Heisenberg ferromagnetic ground state [23]. It would be interesting to extend the latter results to the Boson gas using the WIBG with the c-number substitution as Hamiltonian of the noninteracting quasi-particles.

It is easy to see (see also[8]) that a nonzero critical velocity $v_c$ implies
$$
v_s > 0
\eqno{\rm{(5)}}$$

and therefore a nonzero sound velocity is a necessary condition for the validity of Landau's  criterion. This necessary condition has an analogue for inhomogeneous, i.e., non-translationally invariant systems such as trapped gases.
Let the thermodynamic limit of the (ground state (g.s.), i.e., $T=0$) compressibility
$\,\kappa_0\,$ be defined by
$$
\kappa_0 = {\lim_{V\to\infty}\;
\left[-V^{-1} \left(\dis\frac{\partial V}{\partial P}\right)_{T=0,N} \right]}
\eqno{\rm{(6)}}$$
where
$$
P= - (\partial E (N,L) / \partial V)_{N, T=0}
\eqno{\rm{(7)}}$$
is the g.s. pressure, and $\,E (N,L)\,$ denotes the g.s. energy of
$\, H_N$. For homogeneous systems, and under quite general conditions on the pair interactions $\,v\,$ it may be proved [6] that the limit in (5) exists and
$$
\kappa_0 \geq 0
\eqno{\rm{(8)}}$$
and, by a macroscopic argument [7], one expects that
$\,v_s\,$, given by (4), is alternatively given by
$$
v_s = \left[\frac{1}{m\rho} \kappa_0^{-1}\right]^{1/2}
\eqno{\rm{(9)}}$$
On the other hand, (9) makes sense for general (in particular inhomogeneous) systems such as (2). Comparing (9) with (4), we are led to
adopt as \textbf{necessary part of Landau's criterion for general systems}
the condition:
$$
\kappa_0^{-1} > 0 \eqno{\rm{(10)}}
$$
In order to study (6) for the trapped gases (2) (with $T\ne0$), we 
enclose the system in a large box and should consider the general limit on
the r.h.s. of (6) where $N$, rather than $Na/L$, is fixed, so that we are not
allowed to use the results for the GP limit (1), and the problem is difficult.
One might argue intuitively that the trap potential $V_{T}$ is ``like a box'', and enclosing
the system in a large box whose volume is varied produces no change: thus $P$
is zero in (7), and so is the inverse compressibility, yielding zero for the
r.h.s. of (10). Thus the necessary part of Landau's criterion seems not to 
hold for trapped gases, but for a somewhat trivial reason. This is not so, 
however, because the trap potential is slowly varying, and very different
from the steep ones which resemble a box. In the limit of very slow variation,
one has the \textbf{homogeneous} GP gas which has been thoroughly studied
(see [12],chapter 5). The proofs of Bose Einstein condensation (BEC) and
superfluidity (according to the two-fluid criterion) in the latter reference
have, in fact, been an excelent qualitative guide for the corresponding behavior of trapped gases, and for that reason we turn our attention now to the study of (10) in the homogeneous GP case. 

Let $\frac{N}{V}=\rho$ denote the density. This may be fixed in the homogeneous GP limit but in a trap the particle density is inhomogeneous. We consider two
versions of the GP limit which have been used:
GP1: $Na/L$ and $1/L$ fixed; 
GP2: $\rho$ and $Na/L$ fixed.

 Above, $L$ and $a$ are to be regarded as adimensional (i.e., divided by a length unit. In the trapped case, the natural length unit is the ``trap extension'' $L_osc$, see ([12],p.47). As remarked in ([12], p.40), GP2 may be implemented by replacing the interaction potential $v$ in (2) by
$$
v(x)=a^{-1}v_1(\vert x \vert/a)
\eqno{\rm{(11)}}
$$

with $v_1$ of unit scattering length, keeping $v_1$
fixed. with this choice $a$ tends to zero as
$$
a= N^{-2/3}\rho^{-1/3}
\eqno{\rm{(12)}}
$$
 since 
$$L=(\frac{N}{\rho})^{1/3}\eqno{\rm{(13)}}$$ and (1) holds.
By scaling, the limit involved in the proof of ODLRO (off-diagonal long range order)[20], which implies existence of BEC in a dilute limit, is \textbf{equivalent} to GP1 ([12], Theorem 5.1, pg.40). The same happens with the proof of superfluidity by the two-fluid criterion (3)(see [12],pg.45). This equivalence does not extend, however, to the proof of (10), as we shall see.
 
The g.s. energy in the thermodynamic limit,
$$
e_0 (\rho) = \lim_{L\rightarrow \infty}\limits 
    E_0 (\rho L^3, L ) / ( \rho L^3)
\eqno{\rm{(14)}}$$
is rigorously given in the {\em dilute limit\/} viz,
$$
 \rho a^3 \ll 1 \quad ,
\eqno{\rm{(15)}}$$
,by the seminal result of Lieb and Yngvason [11]:
$$
\lim_{\rho a^3 \rightarrow 0}\limits 
   \frac{e_0 (\rho)}{4\pi \,\,\mu\,\,\rho\,\,a} =1
\eqno{\rm{(6)}}$$
The rigorous results referred to above (for a more complete account, see
[12]) seem, so far, to support the assumption that the Bogoliubov
expansion (in the forthcoming $\alpha \equiv \frac{128}{15 \sqrt{\pi}}$ and
 $\beta \equiv 8(\frac{4\pi}{3} - \sqrt{3})$):
\begin{eqnarray*}
\frac{E_0(N,L,a)/N}{4\pi \mu \rho a} = 1 &+& \\
\alpha(Na^3/V)^{1/2} &+& \\
\beta(Na^3/V)\log (Na^3/V) &+& \\ 
o (Na^3/V) 
\end{eqnarray*}
$$\eqno{(17)}$$
is an asymptotic series in region (15).  We refer to ([12],pg. 11) and Lieb's early review [10c] for details, references and history of (17), but remark that [13] is the simplest, albeit nonrigorous, approach to the derivation of the first two terms of (17).

In the following considerations, the scattering length is a fixed number satisfying (17). In the limit (6) the derivatives with respect to $\, V\,$ are
taken {\em at fixed $N$} and only thereafter the thermodynamic limit, hereafter just denoted by ``lim'', $\, N\rightarrow \infty, \,\, V\rightarrow \infty\,$ with $\,
\dis\frac{N}{V} = \rho\,$ performed. We arrive ,by (6), (7) and (17), at 
\begin{eqnarray*}
\kappa^{-1}_0=\lim \left[V\frac{\partial^2 E_0 (N,L)}
   {\partial V^2}\right]_N = \;
8\pi\mu(\rho^2)a &+& \\
15\pi\mu\alpha (Na)^{5/2}V^{-5/2} &+& \\
8\pi\mu\beta(3N^3 V^{-3} a^4(\log (Na^3/V)+1 )) &+& \\ 
o (Na^3/V^2)
\end{eqnarray*}
$$\eqno{(18)}$$

By (18) and (12) we obtain
$$
\frac{\kappa^{-1}_0}{\kappa^{-1}_{f}}=N^{-2/3}f_2(N)
\eqno{\rm{(19a)}}
$$
where
$$
f_2(N)=1+\frac{15}{8}\alpha N^{-1} + \;
\beta(4N^{-2}-3N^{-2}\log{N})+o(N^{-3})
\eqno{\rm{(19b)}}
$$
Above, $\kappa^{-1}_{f}=8\pi\mu\rho^{5/3}=\frac{\hbar^2}{m}\rho^{5/3}$ denotes the compressibility of free fermions in three dimensions. We assume that $f_2(N)$, defined by (19b), is asymptotic of order two (in the parameter $N^{-1}$) to the function $g=N^{2/3}\frac{\kappa^{-1}_ 0}{\kappa^{-1}_{f}}$: this is a precise version of our assumption regarding asymptoticity. This yields by (19a) and (19b):
$$
\kappa^{-1}_0 =0 \eqno{\rm{(20)}}
$$
and thus the special form (10) of Landau's criterion is not satisfied for
any regime of flow velocities. Thus GP2 implies that the sound velocity is zero by (9).
On the other hand, it is easy to see that (9) and (18) imply that GP1 yields
for the sound velocity the value $1/L$ (up to a constant factor with the dimension of velocity), i.e., a nonzero value!
Both GP1 and GP2 yield a value \textbf{of the same order} for the dilution parameter (15). Taking $L=1$, GP1 yields
$\rho a^3=N\times(1/N)^3=N^{-2}$ (in the trapped case it is the mean density which enters (15), see [12], pg.49) and by GP2 one finds also $\rho a^3=\rho(N^{-2/3})^3=O(N^{-2})$. The values for $E_0/N$ are, however, quite different: for GP1 it is a nonzero constant proportional to $Na$, for GP2 it is , by (15) and (17), proportional to $N^{-2/3}$ or $V^{-2/3}$, which coincides with the behaviour of the \textbf{free} Bose gas with Dirichlet boundary conditions! This qualitative difference in behavior affects the leading term in (17), and therefore the compressibility in (18) but, as we have seen, this does not matter for the presence or absence of superfluidity by the two-fluid criterion (3), since the latter holds even if $E_0=0$.
 
For homogeneous systems, we view GP2 as the most natural, fundamental definition, and GP1 as an equivalent condition for certain quantities. This is also indicated by the fact that the density $\rho$ has no thermodynamic limit by GP1 and is, in fact, the view adopted in [12], in which the GP limit is defined by GP2. From this point of view, Landau's condition should not be valid for the homogeneous GP case, and, we believe, also not for trapped gases in the GP limit, because, as argued above, the presence of the trap seems to strengthen the argument in favor of  a zero value for the compressibility.
 
Condition (10) is, of course, equivalent to finiteness of the
g.s. compressibility and is violated by the isothermal compressibility
of the free Bose gas below the transition temperature (joining smoothly
to $\, T=0$). Clearly, the identity between the two expressions (4) and
(9) is highly nontrivial. To our knowledge, it has been rigorously
proved (by E.H. Lieb [10b])  only for the model of $\, N\, $ particles of mass
$\,m\,$ in one dimension with repulsive delta function interactions [10a] 
whose formal Hamiltonian is given by 
\setcounter{equation}{20}
\begin{eqnarray}
H_N &=& - \frac{\hbar^2}{2m}
    \sum^N_{i=1} \frac{\partial^2}{\partial x^2_i} +
      2c \sum_{\langle i,j\rangle}\limits
     \delta \left(x_i - x_j\right)\nonumber \\[0.3cm]
  && 0\leq x_i \leq L, 1=1, \ldots , N
\end{eqnarray}
where $\, c>0, <i,j>\,$ denote nearest neighbours. 
We now explain why the compressibility of free fermions may be expected to appear in (19a). In the limit $a\to 0$ implied by (12), (14) becomes a hard potential of zero range, just as the limit $\, c\rightarrow \infty\,$ of (21)
(Girardeau's model [14]). This seems to be the only model for which both the (double) spectrum of elementary excitations and the compressibility were obtained rigorously without (as yet) unproved assumptions, the former in [10b]. The forthcoming derivation, which is not explicitly  found in the literature but follows easily from [14], is elementary and provides a transparent physical reason for the validity of property (10). 

In the limit $c\rightarrow\infty$ of (21), the boundary condition on the wave-functions reduces to 
$$
\psi\left(x_1, \ldots, x_N\right) =0 
   \quad \mbox{if} \quad x_j=x_\ell \quad
     1 \leq j<\ell\leq N
\eqno{\rm{(22a)}}
$$
and the (Bose) eigenfunctions satisfying (22.a) simplify to
$$
\psi^B \left(x_1, \ldots, x_N\right) =
    \psi^F \left(x_1, \ldots , x_N\right) 
      A \left(x_1, \ldots , x_N\right)
\eqno{\rm{(22b)}}
$$
where $\, \psi^F\,$ is the Fermi wave function for the free system of
$\,N\,$ particles confined to the region $\, 0\leq x_i < L, \,\,i=1,
\ldots , N$, with periodic boundary conditions: it automatically
satisfies (22a) by the exclusion principle. In (22b),
$$
A\left(x_1, \ldots, x_N\right) = \prod_{j>\ell}\limits 
    sgn \left(x_j - x_\ell\right)
\eqno{\rm{(22c)}}
$$
from the limit $\, c\rightarrow \infty\,$ of [10a], or [14].
From (22b) and the fact that $\, \psi^B_0\,$ is nonnegative (the suffix
zero referring to the ground state), it follows that $\, \psi^F_ 0\,$ has
constant sign in the $\, N!\,$ regions in which configuration space is
divided by the surfaces $\, x_j = x_\ell$, and, thus, from (22.b),
$$
\psi^B_0 = \left| \psi^F_0\right|
\eqno{\rm{(22d)}}
$$
Since $\, A^2 = 1\,$ by (22.c), the correspondence (22.b) preserves all
scalar products, and therefore the energy spectrum of the system is the
same as of the free Fermi gas, in particular the ground state energy
equals (for $\,N\,$ odd): 
$$
E_{0,N} = \frac{\hbar^2}{m} 
    \sum^{\frac{1}{2} (N-1)}_{p=1}\limits
      \left(\frac{2\pi p}{L}\right)^2
\eqno{\rm{(22e)}}
$$
and $\, \psi^F_0\,$ is a Slater determinant of plane-wave functions
labelled by wave-vectors \linebreak $\, k_i, i=1, \ldots , N$, such that
$$
-K(\infty) \leq k_i \leq K(\infty) \qquad i=1,\ldots, N
   \eqno{\rm{(24a)}}
$$
with
$$
K(\infty) = \frac{2\pi}{L} \cdot \frac{1}{2} (N-1) \simeq \pi \rho
  \eqno{\rm{(24b)}}
$$
the region (24a) being the ``Fermi sphere''. A straightforward analysis of the elementary excitations [10b] yields two branches, both corresponding to the same sound velocity by (4): 
$$
v_s = 2K = \pi \rho \,\,\hbar/m 
\eqno{\rm{(25)}}
$$
The r.h.s. of (9) follows by an elementary calculation: from the formula
$$
\sum^N_{k=1}\limits \,\,k^2 = \frac{1}{6} \,\, N (N+1) (2N+1)
$$
we obtain from (8),  and (22e):
$$
-L \frac{\partial p}{\partial L} = 3
    \frac{(\pi \hbar N)^2}{3m} (N-N^{-1}) L^{-3}
\eqno{\rm{(26)}}
$$
By (26), $\,\kappa_0$, given by (6), equals (the limit means
$N\rightarrow\infty$, with $L\rightarrow\infty$ and $N/L = \rho$):
$$
\kappa^{-1}_0 =\lim{\left(-L\frac{\partial p}{\partial L} \right)}\;
    = \frac{\pi^2 \hbar^2}{m} \, \rho^3
\eqno{\rm{(27)}}
$$
and thus the r.h.s. of (9) indeed equals the r.h.s. of (4), by
(25). Thus in one dimension (27) shows that the ratio in (19b)equals one. Although the answer in higher dimensions is not known, it may be expected to be nonzero. Why does this ratio approach zero in the GP limit? We now attempt at an explanation, using the comparison with the homogeneous GP limit.

It is the splitting in momentum space (18b), which is
responsible for (10) in this model, with a $\, K(\infty) \neq
0$, which itself is due to the (non-product) structure of the
g.s. wave function (22b) -- which would be identically zero if $\,
K(\infty) =0!$ In contrast, the latter condition is compatible
with a g.s. wave function which is a {\em product\/} of plane waves in
the {\em same\/} state ($k=0$ for periodic boundary conditions),
as for the free Bose gas, leading to infinite compressibility. 
In the G.P. limit, the n-particle density matrices have a product structure ([12],pg.64), similarly to the free Bose gas, and quite in contrast to the rich structure of the correlation functions of the Girardeau model (see [21] for the case of Dirichlet and Neumann b.c.). 

In ref. [15] we have formulated an alternate criterion applicable to rotational superfluidity, which depends on the property of ODLRO: the ``macroscopic wave-function'' associated to the latter property [20] is independent of the azimutal angle, for a system enclosed in a rotating cylinder. The corresponding physical property is the ``London rigidity'', which is shared by the free Bose gas [16].
In contrast, ODLRO is not required for Landau superfluidity, and, indeed, is proved to be absent for the Girardeau model in [17]. For dilute trapped Bose gases  London rigidity was rigorously proved in [5], being thus one of the very few results in mathematical physics which have been verified experimentally ([18], Fig 7a, pg. 48).

Although translational superfluidity of these gases was also
rigorously proved in [5] using the two-fluid criterion, the necessity part of
Landau's criterion (10) predicts otherwise, under assumptions
on the asymptotic character of the Bogoliubov expansion. Thus question a) posed in the beginning seems to have a negative answer. Concerning question b), the issue can thus only be resolved by experiments, but the latter seem to be, as yet, inconclusive (see [19], Sect. 4.5, for a recent discussion of this point
and references). In particular, if the compressibility is very large (infinite by experimental standards), the two-fluid criterion is the relevant one in case the system is a (translational) superfluid, while the Landau criterion is the relevant one in case no translational superfluidity is found. 

\noindent{\bf Acknowledgement}
We should like to thank A. F. R. de Toledo Piza for illuminating conversations, and the referee for improvements of presentation and content.

\noindent{\bf References}

\begin{list}{}{\setlength{\leftmargin}{7mm}\labelwidth2.5cm
\itemsep0pt \parsep0pt}

\item[{[1]}]{\sc M.H. Anderson, J.R. Ensher, M.R. Matthews, C.E.
								Wieman\/}
								and {\sc
								E.A.Cornell\/}
								--
								Science
								{\bf
								269},
								198
								(1995).

\item[{[2]}] {\sc F. Dalfovo, S, Giorgini, L.P. Pitaevskii\/} and {\sc
S.Stringari\/} -- {\sl Rev. Mod, Phys.\/} {\bf 71}, 463 (1999).
\item[{[3]}] {\sc E. Lieb\/} {\sc R. Seiringer} -- {\sl Phys.  Rev.
								Lett.}
								{\bf
								88},
								170409
								--1--4
								(2002).

\item[{[4]}]{\sc R. Seiringer} -- {\sl J. Phys.\/} {\bf A36}, 9755
(2003).

\item[{[5]}] {\sc E.H. Lieb and R. Seiringer\/} and {\sc J. Yngvason\/} --
{\sl Phys. Rev.\/} {\bf B66}, 134529 (2002).

\item[{[6]}] {\sc M.E. Fisher\/} -- {\sl Arch. Rat. Mech. Anal.\/} {\bf
17}, 377 (1964).

\item[{[7]}] {\sc F.London\/} -- {\sl Superfluids\/} -- Vol II, p.83,
  John Wiley, N.Y.

\item[{[8]}] {\sc L.D. Landau\/} et {\sc E.M. Lifschitz} -- Physique
Statistique - Ed. Mir, Moscou, 1967; {\sc W.F. Wreszinski\/} --
Superfluidity -- Modern Encyclopaedia of Mathemathical Physics, to
appear.

\item[{[9]}] {\sc V.A.Zagrebnov\/} and {\sc J.B.Bru} -- {\sl Phys. Rep.}
{\bf 350}, 291 (2001).

\item[{[10a]}] {\sc E.H. Lieb\/} and {\sc W. Liniger} -- {\sl Phys. Rev.}
{\bf 130}, 1605 (1963).

\item[{[10b]}] {\sc E.H. Lieb\/}-- {\sl Phys. Rev.\/} {\bf 130}, 1616
(1963).

\item[{[10c]}] {\sc E.H. Lieb\/} -- The Bose Fluid, in: W.E. Brittin, ed,
Lecture Notes in Theoretical Physics VII C, University of Colorado
Press, pp. 175--224 (1964).

\item[{[11]}] {\sc E.H. Lieb\/} and {\sc J. Yngvason\/} -- {\sl
Phys. Rev. Lett.\/} {\bf 80}, 2504 (1998).

\item[{[12]}] {\sc E.H. Lieb, R. Seiringer, J.P. Solovej\/} and {\sc
J. Yngvason} -- The Mathematics of the Bose Gas and its Condensation --
Birkh\"auser Verlag -- Basel -- Boston -- Berlin 2005.

\item[{[13]}] {\sc E.H. Lieb\/} -- {\sl Phys. Rev.} {\bf 130}, 2518
(1963).

\item[{[14]}] {\sc M.D. Girardeau\/} -- {\sl J. Math. Phys.\/} {\bf 1},
516 (1960).

\item[{[15]}] {\sc G.L. Sewell\/} and {\sc W.F. Wreszinski\/} -- On the Mathematical theory of Superfluidity-- unpublished.

\item[{[16]}] {\sc J.T. Lewis\/} and {\sc J.V. Pul\`e\/} -- {\sl Comm. Math.
Phys.\/} {\bf 45}, 115 (1975).

\item[{[17]}] {\sc A. Lenard\/} -- {\sl J. Math. Phys.\/} {\bf 5}, 930
								(1964).

\item[{[18]}] {\sc J. Dalibard\/} and {\sc C. Salomon\/} -- S\'eminaire
								Poincar\'e
								{\bf 1},
								39
								(2003).
\item[{[19]}] {\sc M. Brewczyk, M. Gajda\/} and { \sc K. Rzazewski\/}
								--
								{\sl
								J.
								Phys.\/}
								{\bf
								B40},
								R1
								-R37
								(2007),
								section
								4.5
								(pg.
								R30).

\item[{[20]}] {\sc O. Penrose\/} and {\sc L. Onsager\/} -- {\sl Phys. Rev.\/} 
 {\bf 104}, 576 (1956).

\item[{[21]}] {\sc P.J.Forrester, N.E. Frankel\/} and {\sc T.M. Garoni\/} -- 
{\sl J. Math. Phys.\/} {\bf 44}, 4157 (2003).

\item[{[22]}] {\sc W. Wreszinski\/} and {\sc C. Jaekel\/} -- {\sl Ann. Phys.\/}
 {\bf 323}, 251 (2008).

\item[{[23]}] {\sc K. Hepp\/} -- {\sl Phys. Rev.\/} {\bf B5}, 95 (1972).

\item[{[24]}] {\sc Q. Chen\/}, {\sc J. Stajic\/}, {\sc S. Tan\/} and 
{\sc K. Levin\/} -- {\sl Phys. Rep.\/} {\bf 412}, 1 (2005).

\end{list}

\end{document}